\documentclass[10pt,conference]{IEEEtran}

\usepackage{cite}
\usepackage{amsmath,amssymb,amsfonts}
\usepackage{algorithmic}
\usepackage{graphicx}
\usepackage{textcomp}
\usepackage{amsthm}
\usepackage{epstopdf}

\newtheorem{theorem}{Theorem}

\newtheorem{proposition}{Proposition}

\newtheorem{corollary}{Corollary}

\newtheorem{example}{Example}

\usepackage{xcolor}
\def\BibTeX{{\rm B\kern-.05em{\sc i\kern-.025em b}\kern-.08em
    T\kern-.1667em\lower.7ex\hbox{E}\kern-.125emX}}
\begin{document}

\title{Jensen-Shannon Information Based Characterization of the Generalization Error of Learning Algorithms}

\author{\IEEEauthorblockN{Gholamali Aminian \textsuperscript{\textsection}, Laura Toni, Miguel R. D. Rodrigues}
\IEEEauthorblockA{\textit{ Department of Electronic and Electrical Engineering University College London} \\
\{g.aminian, l.toni, m.rodrigues\}@ucl.ac.uk}\\
}

\maketitle
\begingroup\renewcommand\thefootnote{\textsection}
\footnotetext{The first author is supported by the Royal Society Newton International Fellowship, grant no. NIF\textbackslash R1 \textbackslash192656 .}
\endgroup
\begin{abstract}
Generalization error bounds are critical to understanding the performance of machine learning models. In this work, we propose a new information-theoretic based generalization error upper bound applicable to supervised learning scenarios. We show that our general bound can specialize in various previous bounds. We also show that our general bound can be specialized under some conditions to a new bound involving the Jensen-Shannon information between a random variable modelling the set of training samples and another random variable modelling the hypothesis. We also prove that our bound can be tighter than mutual information-based bounds under some conditions.
\end{abstract}
\begin{IEEEkeywords}
Generalization Error Bounds, Mutual Information, Jensen-Shannon Information\end{IEEEkeywords}
\section{Introduction}\label{Sec:Introduction}
Machine learning-based approaches are increasingly adopted to solve various prediction problems in a wide range of applications such as computer vision, speech recognition, speech translation, and many more~\cite{shalev2014understanding},~\cite{bengio2017deep}.
In particular, supervised machine learning approaches learn a predictor -- also known as a hypothesis -- mapping some input variable to an output variable using some algorithm that leverages a series of input--output examples drawn from some underlying (and unknown) distribution \cite{shalev2014understanding}. It is therefore critical to understand the generalization ability of such a predictor, i.e. how the predictor performance on the training set differs from its performance on a testing set (or on the population).

Various approaches have been developed to characterize the generalization error of learning algorithms. These include VC-based bounds~\cite{vapnik1999overview}, algorithmic stability-based bounds ~\cite{bousquet2002stability}, algorithmic robustness-based bounds ~\cite{xu2012robustness}, PAC-Bayesian bounds~\cite{mcallester2003pac}, and many more. However, many of these generalization error bounds cannot explain the generalization abilities of a variety of machine-learning methods for various reasons: some of the bounds depend only on the hypothesis class and not on the learning algorithm; existing bounds do not easily exploit dependencies between different hypotheses; and existing bounds also do not exploit dependences between the learning algorithm input and output.

More recently, approaches leveraging information-theoretic tools have also been emerging to characterize the generalization ability of various learning methods. Such approaches often express the generalization error in terms of certain information measures between the learning algorithm input (the training dataset) and output (the hypothesis), thereby incorporating the various ingredients associated with the learning problem, including the data distribution, the hypothesis space, and the learning algorithm itself.
In particular, building upon pioneering work by Russo and Zou~\cite{russo2019much}, Xu and Raginsky~\cite{xu2017information} have derived generalization error bounds involving the mutual information between the training set and the hypothesis. Bu \textit{et al.}~\cite{bu2019tightening} have derived tighter generalization error bounds involving the mutual information between each individual sample in the training set and the hypothesis. Meanwhile, bounds using chaining mutual information have been proposed in \cite{asadi2018chaining}. Other authors have also constructed information-theoretic based generalization error bounds based on other information measures such as $\alpha$-R\'eyni divergence, $f$-divergence, and maximal leakage~\cite{esposito2019generalization}. Bounds based on Wassertein distances~\cite{lopez2018generalization}, \cite{wang2019information} and bounds based on other divergences~\cite{jiao2017dependence} are also known. Finally, conditional Mutual Information is used to bound the generalization error in \cite{steinke2020reasoning}.

In this work, we also concentrate on the characterization of the generalization ability of (supervised) machine learning algorithms by making a series of contributions:
\begin{enumerate}
    
    \item First, we offer a new approach to bound the (expected) generalization error of learning algorithms based on the use of auxiliary distributions imposed both on the data generation and the hypothesis generation processes.
    
    \item Second, we show that the proposed bounds readily reduce to various existing bounds depending on how one specifies the auxiliary distributions.
    
    \item Third, we also show that one can recover a new generalization error bound expressed via a Jensen-Shannon information measure 
    
    \item Finally, we showcase that our Jensen-Shannon based bounds offer various advantages in relation to mutual information bounds. It is shown that our new bound -- in addition to being always finite -- can also be tighter than existing ones subject to some conditions.
 
\end{enumerate}



It is noteworthy to add -- although the Jensen-Shannon divergence does not appear to have been used to characterize the generalization ability of learning algorithms -- this information-theoretic quantity has been employed to enable some machine learning problems, including adversarial learning~\cite{goodfellow2014generative} and active learning~\cite{melville2005active}. 


We adopt the following notation in the sequel. Upper-case letters denote random variables (e.g., $Z$), lower-case letters denote random variable realizations (e.g. $z$), and calligraphic letters denote sets (e.g. $\mathcal{Z}$). We denote the distribution of the random variable $Z$ by $P_Z$, the joint distribution of two random variables $(Z_1,Z_2)$ by $P_{Z_1,Z_2}$, and the derivative of a real-valued function $f(x)$ with respect to its argument $x$ by $f^{\prime}( \cdot )$. We also adopt throughout the natural logarithm denoted by $\log (\cdot)$.
\section{Problem Formulation}

We consider a standard supervised learning setting where we wish to learn a hypothesis given a set of input-output examples that can then be used to predict a new output given a new input.

In particular, in order to formalize this setting, we model the input data (also known as features) using a random variable $X \in \mathcal{X}$ where $\mathcal{X}$ represents the input set, and we model the output data (also known as predictors or labels) using a random variable $Y \in \mathcal{Y}$ where $\mathcal{Y}$ represents the output set. We also model input-output data pairs using a random variable $Z = (X,Y) \in \mathcal{Z} = \mathcal{X} \times \mathcal{Y}$ where $Z$ is drawn from $\mathcal{Z}$ per some unknown distribution $\mu$. We also let $\mathcal{S} = \{Z_i = (X_i,Y_i)_{i=1}^n\}$ be a training set consisting of a number of input-output data points drawn i.i.d. from $\mathcal{Z}$ according to $\mathcal{\mu}$.

We represent hypotheses using a random variable $W \in \mathcal{W}$ ($W: \mathcal{X} \rightarrow \mathcal{Y}$) where $\mathcal{W}$ is a hypothesis class. Finally, we represent a learning algorithm via a Markov kernel that maps a given training set $\mathcal{S}$ onto an hypothesis $W$ of the hypothesis class $\mathcal{W}$ according to the probability law $P_{W|S}$.


Let us also introduce a (non-negative) loss function $\ell:\mathcal{W} \times \mathcal{Z}  \rightarrow \mathbb{R}^+$ that measures how well a hypothesis predicts an output given an input. We can now define the population risk and the empirical risk associated with a given hypothesis as follows:
\begin{align}
&L_P(w,\mu)\triangleq \int_{\mathcal{Z}}\ell(w,z)\mu(z) dz\\
&L_E(w,\mathcal{S})\triangleq\frac{1}{n}\sum_{i=1}^n \ell(W,Z_i)
\end{align}
respectively. We can also define the (expected) generalization error as follows:
\begin{align}\label{EqMTLEMR}
\overline{\text{gen}}(P_{W|S},\mu)&\triangleq\mathbb{E}_{P_{W,S}}[ L_P(W,\mu)-L_E(W,S)]
\end{align}
This (expected) generalization error quantifies by how much the population risk deviates from the empirical risk. This quantity cannot be computed directly because $\mu$ is unknown, but it can often be (upper) bounded thereby providing a means to gauge the performance of various learning algorithms.

Our goal in the sequel will be to derive (upper) bounds to this generalization error expressed via various information-theoretic measures. In particular, we will use the KL divergence between two distributions $P_X$ and $P_{X^\prime}$ on a common measurable space given by:
$$KL(P_X||P_{X^\prime})\triangleq\int_\mathcal{X}P_X(x)\log\left(\frac{P_X(x)}{P_{X^\prime}(x)}\right)dx $$
We will also use the mutual information and the lautum information between two random variables $X$ and $X'$ with joint distribution $P_{XX'}$ and marginals $P_X$ and $P_{X'}$ given by \cite{palomar2008lautum}:
$$I(X;X^\prime)\triangleq KL(P_{X,X^\prime}||P_X\otimes P_{X^\prime})$$
and
$$L(X;X^\prime)\triangleq KL(P_X\otimes P_{X^\prime}||P_{X,X^\prime})$$
respectively.

Importantly, we will be using the Jensen-Shannon divergence between two distributions $P_X$ and $P_{X^\prime}$ given by~\cite{lin1991divergence}:
\begin{align}\label{Eq: Jensen-Shannon Definition}
    &\textit{JS}(P_X,P_{X^\prime})\triangleq\\\nonumber
    &\frac{1}{2} \cdot KL\left(P_X\Big|\Big|\frac{P_X+P_{X^\prime}}{2}\right)+\frac{1}{2}KL\left(P_{X^\prime}\Big|\Big|\frac{P_X+P_{X^\prime}}{2}\right)
\end{align}


We will also be using the Jensen-Shannon information between two random variables $X$ and $X'$ with joint distribution $P_{XX'}$ and marginals $P_X$ and $P_{X'}$ given by~ \cite{sason2016f}:
\begin{align}
I_{\textit{JS}}(X;X^\prime)\triangleq\textit{JS}(P_{X,X^\prime},P_X\otimes P_{X^\prime})
\end{align}

It has been shown that the Jensen-Shannon divergence obeys $0\leq \textit{JS}(P_{X},P_{X^\prime}) \leq \log(2)$~ \cite{lin1991divergence} and that Jensen-Shannon information -- which is symmetric -- is zero if and only if the random variables $X$ and $X^\prime$ are independent.

Another quantity that we will often resort to in the sequel relates to the total variation distance between two probability distributions $P_X$ and $P_{X^\prime}$ given by
\begin{align}
    TV(P_X,P_{X^\prime})\triangleq\int |P_X(x)-P_{X^\prime}(x)|dx
\end{align}

\section{Main Results}\label{Sec:ExpectedGEUpper}

We now offer a series of bounds to the expected generalization error based on different information measures. Our analysis also relies on the cumulant generating function of a random variable $X$ given by:
\begin{align}
    \Lambda_X(\lambda)\triangleq\log\left(\mathbb{E}[e^{\lambda(X-\mathbb{E}[X])}]\right)
\end{align}
In Theorem \ref{Theorem: General Upper Bounds} and Theorem \ref{Theorem: Avergae Upper bound}, the bounds to expected generalization error and absolute value of the expected generalization error are presented, respectively.
\begin{theorem}\label{Theorem: General Upper Bounds}
Let us assume that -- under an auxiliary joint distribution $\widehat{P}_{W,Z_i}$ -- $\Lambda_{\ell(W,Z_i)}(\lambda)$ exists, it is bounded by $\psi_+(\lambda)$ for $\lambda \in [0,b_+)$, $0<b_+<+\infty$, and it is also bounded by $\psi_-(-\lambda)$ for $\lambda \in (b_-,0]$, $\forall i=1,\cdots,n$. Let us also assume that $\psi_+(\lambda)$ and $\psi_-(\lambda)$ are convex functions and $\psi_-(0)=\psi_+(0)=\psi_+^\prime(0)=\psi_-^\prime(0)=0$. Then, it holds that:
\begin{align}\label{Eq: Theorem: General Upper Bounds positive}
    \overline{\text{gen}}(P_{W|S},\mu)&\leq \frac{1}{n} \sum_{i=1}^n \left(\psi_+^{\star -1}(A_i)+\psi_-^{\star -1}(B_i)\right)\\
    -\overline{\text{gen}}(P_{W|S},\mu)&\leq \frac{1}{n}  \sum_{i=1}^n \left(\psi_-^{\star -1}(A_i)+\psi_+^{\star -1}(B_i)\right)
\end{align}
where $A_i=KL(P_W\otimes\mu||\widehat{P}_{W,Z_i})$, $B_i=KL(P_{W,Z_i}||\widehat{P}_{W,Z_i})$, \begin{align*}
&\psi_-^{\star -1}(x)=\inf_{\lambda \in [0,-b_-)}\frac{x+\psi_-(\lambda)}{\lambda}
\end{align*}
and
\begin{align*}
&\psi_+^{\star -1}(x)=\inf_{\lambda \in [0,b_+)}\frac{x+\psi_+(\lambda)}{\lambda}\,.
\end{align*} 
\end{theorem}

\begin{theorem}\label{Theorem: Avergae Upper bound}
Let us assume that the loss function is $\sigma$-subgaussian \footnote{A random variable $X$ is $\sigma$-subgaussian if $E[e^{\lambda(X-E[X])}]\leq e^{\frac{\lambda^2 \sigma^2}{2}}$ for all $\lambda \in \mathbb{R}$.} -- under the distribution $\widehat{P}_{W,Z_i}$ $\forall i=1,\cdots,n$-- Then, it holds that:
\begin{align}\label{Eq: Proposition: General Upper Bounds }
    &|\overline{\text{gen}}(P_{W|S},\mu)|\leq \frac{2}{n}  \sum_{i=1}^n \sqrt{\sigma^2\left(A_i+B_i\right)}
\end{align}
where $A_i=KL(P_W\otimes\mu||\widehat{P}_{W,Z_i})$ and $B_i=KL(P_{W,Z_i}||\widehat{P}_{W,Z_i})$\,.
\end{theorem}
Theorem 1 can be used to recover old and new generalization error bounds. For example, we can immediately recover from Theorem 1 the following results.

\begin{example}\label{example: other bounds}
 Let us choose $\widehat{P}_{W,Z_i} = P_W\otimes\mu$ for $i=1,\cdots,n$. It follows immediately from Theorem \ref{Theorem: General Upper Bounds} that:
       \begin{align}\label{Eq:Corollary: other bounds mutual information}
          \overline{\text{gen}}(P_{W|S},\mu)&\leq \frac{1}{n} \sum_{i=1}^n \psi_-^{\star -1}(I(W;Z_i))\\
          -\overline{\text{gen}}(P_{W|S},\mu)&\leq \frac{1}{n} \sum_{i=1}^n \psi_+^{\star -1}(I(W;Z_i))
        \end{align} 
\end{example}  
\begin{example}
Let us now choose $\widehat{P}_{W,Z_i} = P_{W,Z_i}$ for $i=1,\cdots,n$. It also follows immediately from Theorem \ref{Theorem: General Upper Bounds} that:
\vspace{-15pt}
         \begin{align}\label{Eq:Corollary: other bounds Lautum information}
          \overline{\text{gen}}(P_{W|S},\mu)&\leq \frac{1}{n} \sum_{i=1}^n \psi_+^{\star -1}(L(W;Z_i))\\
          -\overline{\text{gen}}(P_{W|S},\mu)&\leq \frac{1}{n} \sum_{i=1}^n \psi_-^{\star -1}(L(W;Z_i))
        \end{align} 
\end{example}
The result in Example 1 corresponds to a result appearing in~\cite{bu2019tightening} whereas the result in Example 2 is a generalization of the result appearing in \cite{gastpar2019Lautum}. It is also worthwhile to mention that the auxiliary joint distributions are different in Examples 1 and 2. Therefore, the $\psi_+$ and $\psi_-$ functions are different for different auxiliary joint distributions. 

Importantly, we can also use Theorem 2 to offer a new generalization error bound based on the Jensen-Shannon information.
\begin{corollary}\label{corollary: Subgaussian and Jensen}
Assume that the loss function $\ell(W,Z_i)$ is $\sigma$-subgaussian under the distribution $\widehat{P}_{W,Z_i}=\frac{P_{W,Z_i}+P_W\otimes\mu}{2}$, $\forall i=1,\ldots,n$. 
It then follows that:
\begin{align}\label{Eq: corolary subgaussian result}
    |\overline{\text{gen}}(P_{W|S},\mu)|\leq \frac{2}{n}\sum_{i=1}^n\sqrt{2\sigma^2I_{\textit{JS}}(W;Z_i)}
\end{align}
\end{corollary}

Note that this result also applies immediately to any bounded loss function $l:\mathcal{W}\times \mathcal{Z}\rightarrow [a,b]$ in view of the fact that such functions are $(\frac{b-a}{2})$-subgaussian under all distributions, \cite{xu2017information}. Note also that this result cannot be immediately recovered from existing approaches such as~\cite[Theorem.~2]{esposito2019generalization}. Notably, there exist $f$-divergence based representations of the Jensen-Shannon information as follows:
\begin{align}
    \textit{JS}(P_X,P_{X^\prime})=\int P_{X}(x) f\left(\frac{P_{X^\prime}(x)}{P_X(x)}\right)dx 
\end{align}
with $f(t)=t\log(t)-(1+t)\log(\frac{1+t}{2})+\log(2)$. However, \cite[Theorem.~2]{esposito2019generalization} requires that the function $f(t)$ associated with the $f$-divergence is non-decreasing within the interval $[0,+\infty)$, but such a requirement is naturally violated by the function $f(t)=t\log(t)-(1+t)\log(\frac{1+t}{2})+\log(2)$ associated with the Jensen-Shannon divergence.

The value of this new proposed bound displayed in Corollary \ref{corollary: Subgaussian and Jensen} in relation to existing bounds can also be further appreciated by offering two additional results.

\begin{proposition}\label{remark: Bounded JS upper} Consider the assumptions in Corollary \ref{corollary: Subgaussian and Jensen}. Then, it follows that:
\begin{align}\label{Eq: proposition: Bounded JS upper}
    |\overline{\text{gen}}(P_{W|S},\mu)|\leq 2\sigma \sqrt{2\log(2)}\approx 2.3548 \sigma
\end{align}
\end{proposition}
This proposition showcases that the proposed Jensen-Shannon divergence generalization bound is always finite, in sharp contrast with existing mutual information and lautum information based bounds that do not have to be bounded (e.g. \cite{xu2017information}, \cite{bu2019tightening}, \cite{asadi2018chaining} and \cite{esposito2019generalization}). This result applies independently of whether or not the loss function is bounded.\footnote{Naturally, it is possible to show that the absolute value of the expected generalization error is always upper bounded as follows $|\overline{\text{gen}}(P_{W|S},\mu)|\leq (b-a)$ for any bounded loss function within the interval $[a,b]$.}

The next proposition instead showcases that the Jensen-Shannon information bound can be lower than the mutual information based bound under certain conditions.

\begin{proposition}
Consider the assumptions in Corollary \ref{corollary: Subgaussian and Jensen}. Then, it follows that the Jensen-Shannon information upper bound given by:
 \vspace{-10pt}
\begin{equation}\label{Eq: JS upper bound}
    |\overline{\text{gen}}(P_{W|S},\mu)| \leq \frac{2}{n} \sum_{i=1}^n \sqrt{2\sigma^2 I_{\textit{JS}}(W;Z_i)}
\end{equation}
is lower than the mutual information based upper bound,\cite{bu2019tightening}, given by:
 \vspace{-10pt}
\begin{equation}\label{Eq: MI uper bound}
    |\overline{\text{gen}}(P_{W|S},\mu)| \leq \frac{1}{n} \sum_{i=1}^n \sqrt{2\sigma^2I(W;Z_i)}
\end{equation}
provided that $8 \log(2)^2\leq I(W;Z_i)$ holds for $i=1,\cdots,n$.
\end{proposition}

\section{Proofs}

\subsection{Proof of Theorem 1}
The proofs of the bounds to $\overline{\text{gen}}(P_{W|S},\mu)$ and $-\overline{\text{gen}}(P_{W|S},\mu)$ are similar. We therefore focus on the later. 


Let us consider the variational representation of KL divergence between two probability distributions $\alpha$ and $\beta$ on a common space $\Psi$ given by~\cite{dupuis2011weak}:
\begin{align}
    KL(\alpha||\beta)=\sup_{f} \int_\Psi f d\alpha -\log\int_\Psi e^f d\beta
\end{align}
where $f\in \mathcal{F}=\{f:\Psi\rightarrow\mathbb{R} \text{ s.t.  } \mathbb{E}_{\beta}[e^f]< \infty \}$.

We can now use the variational representation to bound $KL(P_{W,Z_i}||\widehat{P}_{W,Z_i})$ for $\lambda \in (b_-,0]$ as follows:
\begin{align}\label{Eq:Theorem 1 GE KL Joint}
    &KL(P_{W,Z_i}||\widehat{P}_{W,Z_i})\geq \\\nonumber &\mathbb{E}_{P_{W,Z_i}}[\lambda \ell(W,Z_i)]-\log \mathbb{E}_{\widehat{P}_{W,Z_i}}[e^{\lambda \ell(W,Z_i)}]\geq\\\label{Eq: Theorem 1 GE KL CGF bound}
    &\lambda(\mathbb{E}_{P_{W,Z_i}}[ \ell(W,Z_i)]- \mathbb{E}_{\widehat{P}_{W,Z_i}}[\ell(W,Z_i)])-\psi_-(-\lambda)
\end{align}
where the last inequality is due to:
\begin{align}
  &\Lambda_{\ell({W},{Z}_i)}(\lambda)=\\\nonumber
  &\log\left(\mathbb{E}_{\widehat{P}_{W,Z_i}}[e^{\ell(W,Z_i)-\mathbb{E}_{\widehat{P}_{W,Z_i}}[\ell(W,Z_i)]}]\right)\leq \psi_-(-\lambda)
\end{align}
 It can then be shown from \eqref{Eq: Theorem 1 GE KL CGF bound} that following holds for $\lambda \in (b_-,0]$:
\begin{align}\label{Eq: GE upper BU 2}
    &\mathbb{E}_{\widehat{P}_{W,Z_i}}[\ell(W,Z_i)] - \mathbb{E}_{{P}_{W,Z_i}}[ \ell(W,Z_i)]\leq\\\nonumber
    &\inf_{\lambda \in [0,-b_-)} \frac{B_i+\psi_-(\lambda)}{\lambda}=
    \psi_-^{\star -1}(B_i)
\end{align}
where $B_i=KL(P_{W,Z_i}||\widehat{P}_{W,Z_i})$.
It can likewise also be shown by adopting similar steps that the following holds for $\lambda \in [0,b_+)$:
\begin{align}
    &\mathbb{E}_{P_{W,Z_i}}[ \ell(W,Z_i)]-\mathbb{E}_{\widehat{P}_{W,Z_i}}[\ell(\widehat{W},\widehat{Z}_i)]\leq\\\nonumber
    &\inf_{\lambda \in [0,b_+)} \frac{B_i+\psi(\lambda)}{\lambda}=
    \psi_+^{\star -1}(B_i)
\end{align}
We can similarly show using an identical procedure that:
\begin{align}\label{Eq: GE upper BU 1}
    &\mathbb{E}_{P_{W}\otimes \mu}[\ell({W},{Z}_i)]- \mathbb{E}_{\widehat{P}_{W,Z_i}}[\ell(\widehat{W},\widehat{Z}_i)] \nonumber\leq
    \psi_+^{\star -1}(A_i)\\
    &\mathbb{E}_{\widehat{P}_{W,Z_i}}[\ell(\widehat{W},\widehat{Z}_i)]-\mathbb{E}_{P_{W}\otimes \mu}[ \ell({W},{Z}_i)]\nonumber\leq\psi_-^{\star -1}(A_i)
\end{align}
Where $A_i=KL(P_W\otimes\mu||\widehat{P}_{W,Z_i})$.

Finally, we can immediately bound the generalization error by leveraging \eqref{Eq: GE upper BU 1} and \eqref{Eq: GE upper BU 2} as follows:
\begin{align*}
   &\overline{\text{gen}}(P_{W|S},\mu)=
   \frac{1}{n}\sum_{i=1}^n\mathbb{E}_{P_{W}\otimes \mu}[\ell({W},{Z}_i)]-\mathbb{E}_{{P}_{W,Z_i}}[\ell(W,Z_i)]\\
   &=\frac{1}{n}\sum_{i=1}^n \mathbb{E}_{P_{W}\otimes \mu}[\ell({W},{Z}_i)]- \mathbb{E}_{\widehat{P}_{W,Z_i}}[\ell({W},{Z}_i)]
    +\\
    &\quad~\mathbb{E}_{\widehat{P}_{W,Z_i}}[\ell({W},{Z}_i)]-\mathbb{E}_{{P}_{W,Z_i}}[ \ell(W,Z_i)]\\
   & \leq\frac{1}{n}\sum_{i=1}^n ((\psi_+^{\star -1}KL(P_W\otimes\mu||\widehat{P}_{W,Z_i}))\\&\quad+\psi_-^{\star -1}(KL(P_{W,Z_i}||\widehat{P}_{W,Z_i})))
\end{align*}
\subsection{Proof of Theorem 2}
The assumption that the loss function is $\sigma$-subgaussian under the distribution $\widehat{P}_{W,Z_i}$ implies that $\psi_-^{\star -1}(y)=\psi_+^{\star -1}(y)=\sqrt{2\sigma^2y}$, \cite{bu2019tightening}. It then follows immediately that:
\begin{align}
    |\overline{\text{gen}}(P_{W|S},\mu)| & \leq \frac{1}{n} \sum_{i=1}^n \left(\sqrt{2\sigma^2 A_i}+\sqrt{2\sigma^2 B_i}\right)\\
    &\leq \frac{2}{n} \sum_{i=1}^n \sqrt{\sigma^2(A_i+B_i)}
\end{align}
where the last inequality follows from the concavity of function $\sqrt{2\sigma^2y}$.

\subsection{Proof of Corollary 1}
This result follows immediately from the Jensen-Shannon information definition \eqref{Eq: Jensen-Shannon Definition} by using $\widehat{P}_{W,Z_i}=\frac{P_{W,Z_i}+P_W\otimes\mu}{2}$ in Theorem 2.


\subsection{Proof of Proposition 1}
This follows immediately from \eqref{Eq: corolary subgaussian result} in view of the fact that $I_\textit{JS}(W,Z_i))\leq \log(2)$ for $i=1,\cdots,n$.

\subsection{Proof of Proposition 2}
We start by showing that 
\begin{align}\label{Eq: final result for compare}
    I^2_{\textit{JS}}(X;Y)\leq \frac{\log(2)^2I(X;Y)}{2}
\end{align}
This can be proved immediately by combining the well-known Pinsker's inequality given by:
\begin{align}\label{Eq: Pinsker inequality}
    TV(P_{X,Y},P_X\otimes P_Y)\leq \sqrt{2I(X;Y)}
\end{align}
with another results given by~\cite{lin1991divergence}:
\begin{align}\label{Eq: JS TV compare}
    2I_{\textit{JS}}(X;Y)\leq \log(2) TV(P_{X,Y},P_X\otimes P_Y)
\end{align}


Therefore, we can now prove the proposition by showing that $$4I_{\textit{JS}}(W;Z_i)\leq I(W;Z_i)$$
holds whenever $8\log(2)^2 \leq I(W;Z_i)$.
First, notice that it follows from \eqref{Eq: final result for compare} that:
\begin{align}\label{Eq: first}
&4I_{\textit{JS}}(W;Z_i)\leq 2\log(2)\sqrt{2}\sqrt{I(W;Z_i)}
\end{align}
Next, notice that the equality
$$2\log(2)\sqrt{2}\sqrt{I(W;Z_i)}-I(W;Z_i)=0$$ has two roots corresponding to $I(W;Z_i)=0$ and $I(W;Z_i)=8 \log(2)^2$.
Finally, note also that it holds that:
\begin{align}
    2\log(2)\sqrt{2}\sqrt{I(W;Z_i)}\leq I(W;Z_i)
\end{align}
whenever $8 \log(2)^2\leq I(W;Z_i)$. This completes the proof.
\section{Numerical Example}
We now illustrate that our proposed bounds can be tighter than existing ones in a simple setting involving the estimation of the mean of a Gaussian random variable $Z \sim \mathcal{N}(\alpha,\sigma^2)$ based on two i.i.d. samples $Z_1$ and $Z_2$. We consider the hypothesis (estimate) given by $W=tZ_1+(1-t)Z_2$ for $0<t<1$. We also consider the loss function given by:
\begin{align}
  \ell(w,z)=
  \begin{cases}
  (w-z)^2, \quad &\text{if } |w-z|\leq c\\
  c^2, \quad &\text{otherwise}
  \end{cases}
\end{align}

In view of the fact that the loss function is bounded within the interval $[0,c^2]$, it is also $\frac{c^2}{2}$-subgaussian so that we can apply the generalization error upper bounds \eqref{Eq: MI uper bound} and \eqref{Eq: JS upper bound} 
as follows:
\begin{align}\label{Eq: upper MI bound Sim}
    &\overline{\text{gen}}(P_{W|Z_1,Z_2},P_Z)\leq \frac{c^2}{4}\left(\sqrt{2I(W;Z_1)}+\sqrt{2I(W;Z_2)}\right)\\\label{Eq: upper JS bound Sim}
    &\overline{\text{gen}}(P_{W|Z_1,Z_2},P_Z)\leq \frac{c^2}{2}\left(\sqrt{2I_{\textit{JS}}(W;Z_1)}+\sqrt{2I_{\textit{JS}}(W;Z_2)}\right)
\end{align}

It can be immediately shown that $W \sim \mathcal{N}(\alpha,\sigma^2(t^2+(1-t)^2))$ and $(W,Z_1)$ and $(W,Z_2)$ are jointly Gaussian with correlation coefficients $\rho_1=\frac{t}{\sqrt{t^2+(1-t)^2}}$ and $\rho_2=\frac{(1-t)}{\sqrt{t^2+(1-t)^2}}$. Therefore, it also be shown that the mutual informations appearing above are given by $I(W;Z_1)=-\frac{1}{2}\log(1-\rho_1^2)$ and $I(W;Z_2)=-\frac{1}{2}\log(1-\rho_2^2)$. In contrast, the Jensen-Shannon information appearing above can be computed via an entropic based formulation of this information measure as follows \cite{lin1991divergence}:
\begin{align}
    &I_{\textit{JS}}(W;Z_i)=\\\nonumber
    &h\left(\frac{P_{W,Z_i}+P_W\otimes P_{Z_i}}{2}\right)-\frac{1}{2}(h(P_W)+h(P_{Z_i})+h(P_{Z_i,W}))
\end{align}
-- with $h (\cdot)$ denoting the differential entropy -- where
\begin{align*}
 &h(P_{Z_i})=\frac{1}{2}\log(2\pi\sigma^2e),\\
  &h(P_W)=\frac{1}{2}\log(2\pi\sigma^2(t^2+(1-t)^2)e),\\
  &h(P_{W,Z_i})=\log(2\pi\sigma^2e(t^2+(1-t)^2)(1-\rho_i^2)),
\end{align*}
 whereas $h(\frac{P_{W,Z_i}+P_W\otimes P_{Z_i}}{2})$ can be computed numerically.

\begin{figure}
    \centering
    \includegraphics[scale=0.4]{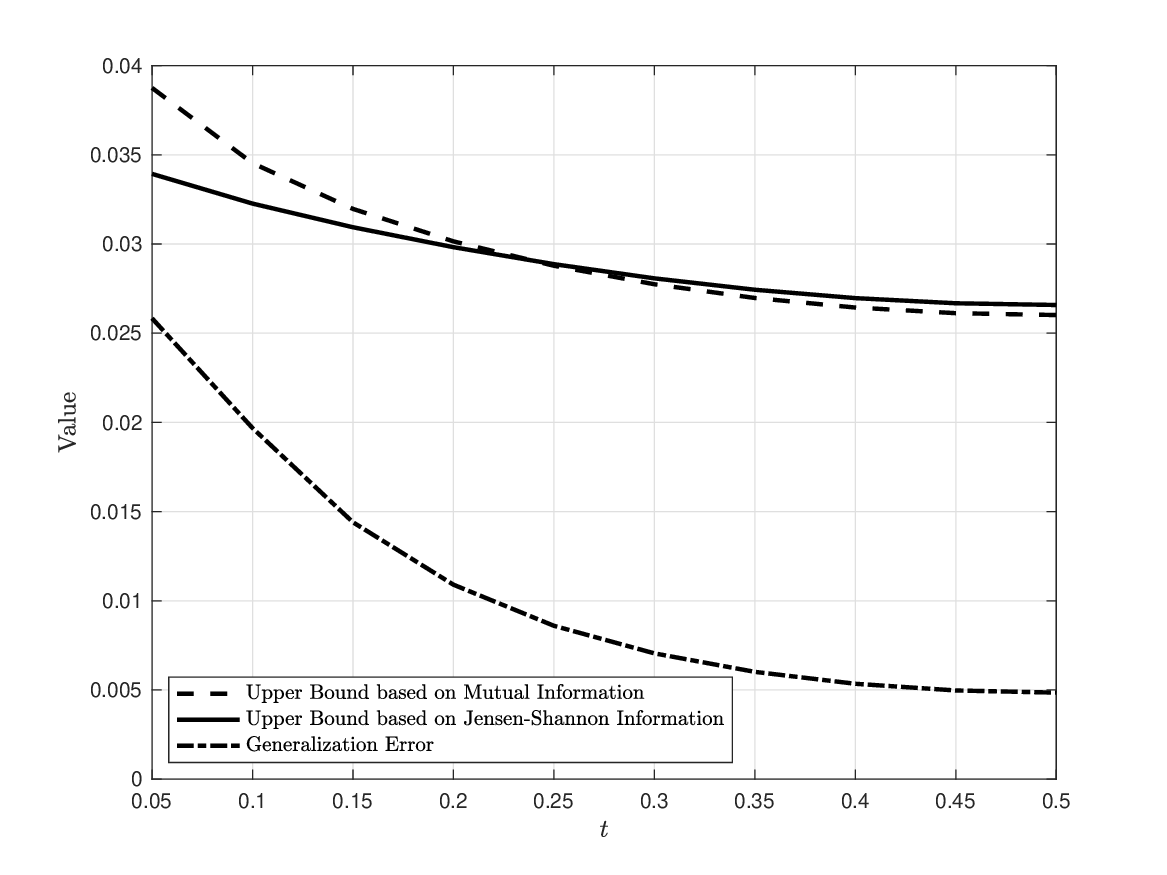}
    \caption{True generalization error, Jensen-Shannon based bound, and Mutual Information based bound.}
    \label{fig:Upper bounds compare GE sigma 1}
\end{figure}

Fig.\ref{fig:Upper bounds compare GE sigma 1} depicts the true generalization error, the mutual information based bound in \eqref{Eq: upper MI bound Sim}, and the Jensen-Shannon information based bound in \eqref{Eq: upper JS bound Sim} for values of $t\in(0,0.5]$, considering $\sigma^2=1, 10$, $\mu=1$, $c=\frac{\sigma}{4}$. It can be seen that for $t<0.25$ the Jensen-Shannon information bound is tighter than the mutual information bound; in contrast, for $t>0.25$, the mutual information bound is slightly better than the Jensen-Shannon information bound. This showcases indeed that our proposed bounds can be tighter than existing ones in some regimes.


\section{Conclusion}\label{Conc}
We have introduced a new approach to obtain information-theoretic oriented bounds to the generalization error associated with supervised learning problems. Our approach can be used to recover organically existing bounds but also to derive new ones based on Jensen-Shannon information measures. Notably, it is shown that the new Jensen-Shannon information can be tighter in some regimes in comparison to existing bounds.
\bibliographystyle{ieeetr}
\bibliography{Refs}
\end{document}